\documentclass[11pt,epsfig]{article}
\usepackage{epsfig}
\usepackage{amsmath}
\usepackage{natbib}
\title{{\bf Spectrophotometric Evolutionary Models\\for Tidal Dwarfs}}
\author{Peter~M.~Weilbacher, Uta~Fritze-v.Alvensleben\\
\vspace{0.1cm}\\
\normalsize Universit\"ats-Sternwarte G\"ottingen, Geismarlandstr.~11, D-37083 G\"ottingen, Germany\\
}
\date{}
\begin{document}
\maketitle
\def\bull{\vrule height .9ex width .8ex depth -.1ex}
\makeatletter
\def\ps@plain{\let\@mkboth\gobbletwo
\def\@oddhead{}\def\@oddfoot{\hfil\tiny
``Dwarf Galaxies and their Environment'';
International Conference in Bad Honnef, Germany, 23-27 January 2001}%
\def\@evenhead{}\let\@evenfoot\@oddfoot}
\makeatother

\begin{abstract}\noindent
  We describe our procedure to model Tidal Dwarf Galaxies in the
  context of our evolutionary synthesis code. Our analysis shows
  strong contributions of gaseous emission lines to optical colors and
  of continuum emission to NIR colors during the burst phase. This
  underlines the importance of including both types of emission when
  modeling any type of star-bursting galaxies.
\end{abstract}

\section{Introduction}

Tidal Dwarf Galaxies (short \textbf{TDGs}) are formed of stars and gas
from (outer) parts of spiral galaxies ripped out from the disk(s)
during interactions and merging events
\citep[see][]{DM94,DBW+97,DBS+00}.  If they get tightly bound they may
remain stable and survive as independent dwarf galaxies, if they do
not fall back onto the interacting system.

Due to the current formation scenario (see Weilbacher \& Duc, this
volume) TDGs can contain both old and young stellar components in
various proportions, inherited from the progenitor spiral disk and
created in a starburst within the tidal tail, respectively. See
\citet{DM98} and \citet{DBS+00} for the most extreme cases. A special
star formation history is therefore required to appropriately model
TDGs.

Spectrophotometric modelling of these objects will help us to
understand their further evolution:\\[-20pt]
\begin{itemize}
\item How bright can they get?\\[-20pt]
\item How strong are the starbursts?\\[-20pt]
\item How much do they fade?\\[-20pt]
\end{itemize}

\section{Model Implementation}

Ingredients of our evolutionary synthesis models \citep[see][for
details]{WDF+00} are Geneva stellar tracks in 3 different
metallicities Z1=0.001, Z2=0.004, Z3=0.008$\approx$Z$_\odot$/2 and the
Scalo IMF.  To approximate a TDG-specific star formation history we
include an underlying old component with constant SFR for the
progenitor disk and a young component due to a starburst of varying
duration and intensity which can be modeled as instantaneous,
exponentially declining or Gaussian-shaped burst (possibly more
realistic). Each model then defines a burst strength
\[b = \dfrac{\mathrm{stellar\ mass\ from\ burst}}{\mathrm{total\ stellar\ mass}}\]

\begin{figure}[t]
  \begin{center}
    \begin{minipage}[c]{0.5\linewidth}
      \epsfig{file=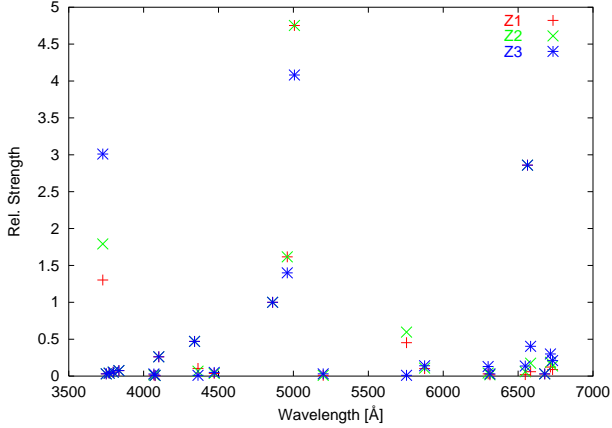,width=\linewidth}
    \end{minipage}
    \parbox[c]{0.45\linewidth}{
      \caption{Line ratios normalized to H$_\beta$ for our three metallicities.}
    }
    \label{fig:lines}
  \end{center}
\end{figure}

To include gaseous emission into our code we use Lyman continuum
photons \citep{SdK97}, gas continuum emission, and emission lines
appropriate for each metallicity. The emission line ratios for the
three metallicities are taken from spectra observed by
\citet{ITL94,ITL97} and \citet{IT98} and theoretical modelling
\citep{Sta84} for low metallicity (Z1, Z2) and medium metallicity
(Z3), respectively. The line ratios we extracted from these works are
displayed in Fig.~\ref{fig:lines}.

\subsection{Broad Band Colors}

\begin{figure}[b]
  \begin{center}
    \epsfig{file=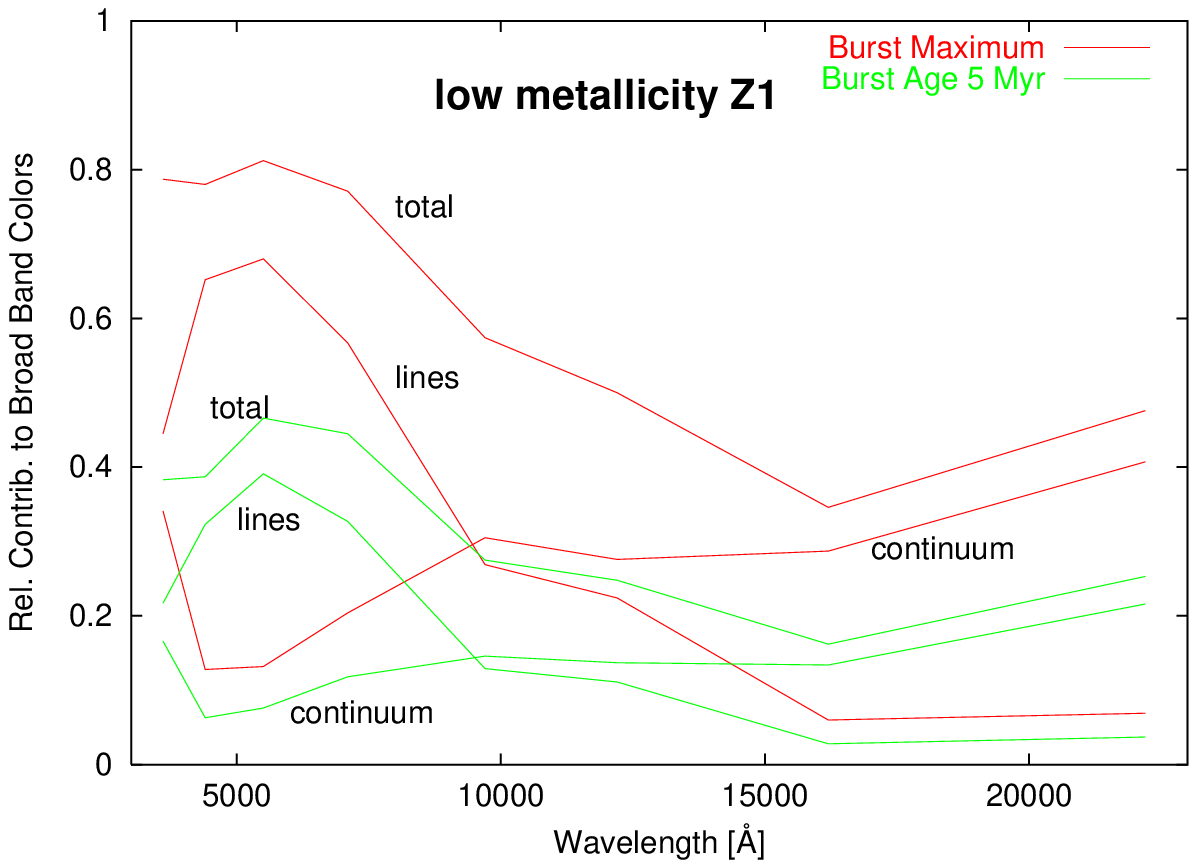,width=0.45\linewidth}
    \epsfig{file=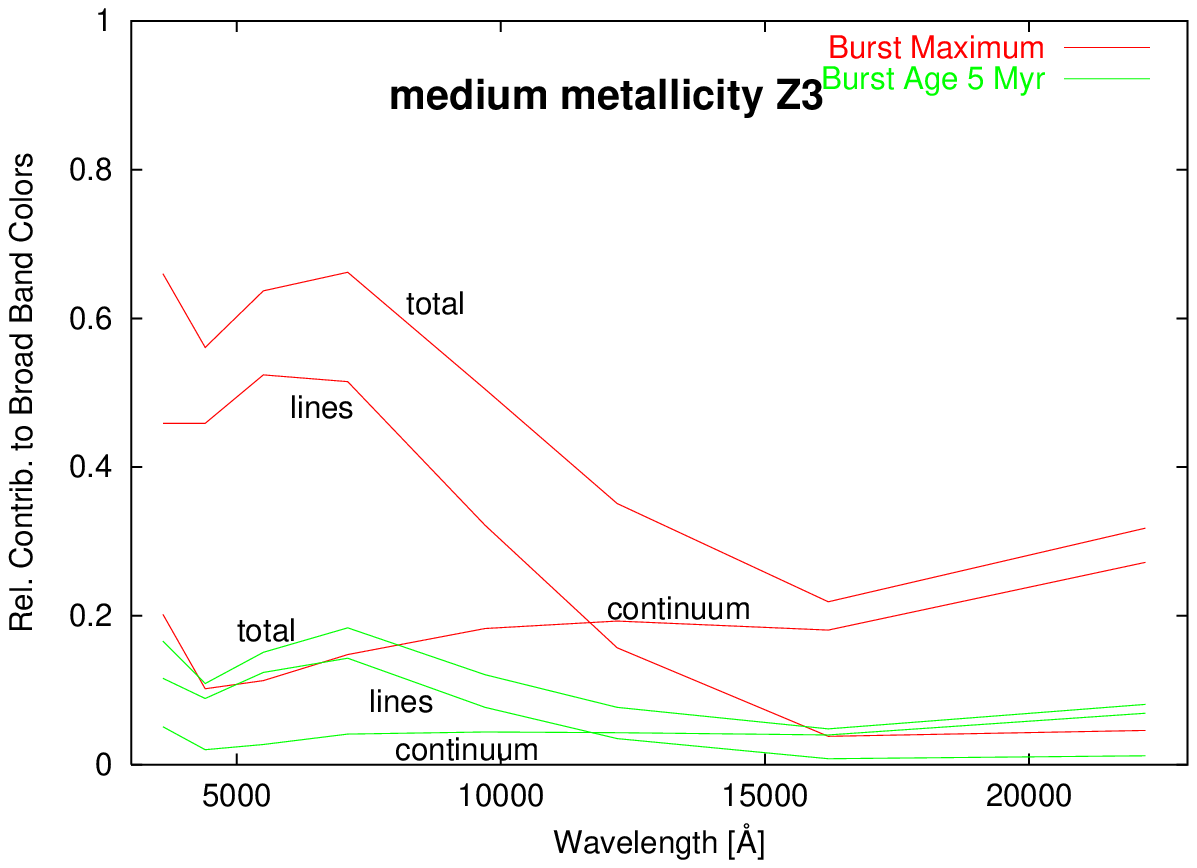,width=0.45\linewidth}
    \caption{Contribution of the gaseous continuum and line emission for 
             two metallicities and two burst ages, $b=0.18$.}
    \label{fig:gas}
  \end{center}
\end{figure}

The contribution of emission lines and gaseous continuum to the
broad-band colors varies with burst age, strength, and metallicity and
strongly depends on the wavelength of the filter, as shown in
Fig.~\ref{fig:gas} for two different ages within the same burst and
two different metallicities.

It is obvious that during the beginning of the burst the optical
colors are dominated by the gaseous emission and that the stellar
contributions only play a minor role, while they gain importance a few
Myrs after the maximum of the burst. The contributions of emission
lines and continuum to the broad-band filters therefore have to be
taken into account.

\begin{figure}[t]
  \begin{center}
    \begin{minipage}[c]{0.5\linewidth}
      \epsfig{file=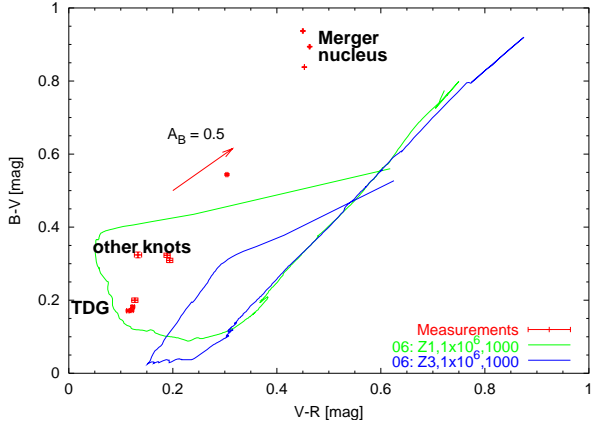,width=\linewidth}
    \end{minipage}
    \parbox[c]{0.45\linewidth}{
      \caption{Two-color diagram $B-V$ vs.~$V-R$ for two metallicities 
        with example data points.}
    }
    \label{fig:broad}
    \vspace*{-20pt}
  \end{center}
\end{figure}

Using two-color diagrams we can now compare the photometry of
observations and models, to derive burst age, burst strength, and
further evolution. In the case shown in Fig.~\ref{fig:broad} the model
with Z1 is fitting better than the model with Z3. For a detailed
discussion of this type of diagram see \citet{WDF+00}. If photometry
in more than three filters is available, the results will be more
accurate, especially when including NIR colors.

\subsection{Spectra}

To better constrain the parameters, obtain information about internal
extinction, and disentangle ambiguities from metallicity, burst age,
and burst strength, we also want to compare with \emph{spectra} of the
objects.  We therefore create model spectra at several timesteps
during and after the burst, using the stellar library from
\citet{LCB97,LCB98}.  We had to interpolate the spectra linearly to 2
\AA\ resolution in the optical to properly add Gaussian-shaped
emission lines with typical FWHM onto the synthetic galaxy spectra.

\begin{figure}[b]
  \begin{center}
    \epsfig{file=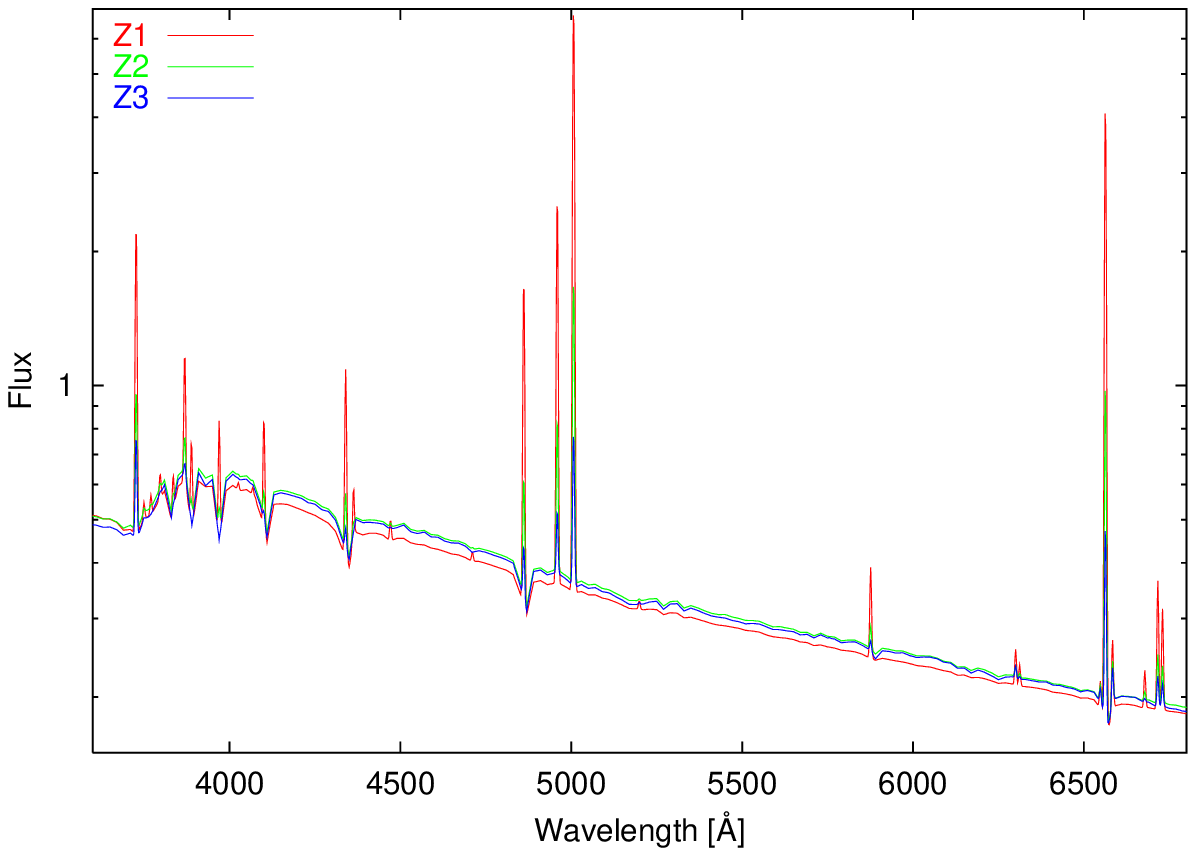,width=0.45\linewidth}
    \epsfig{file=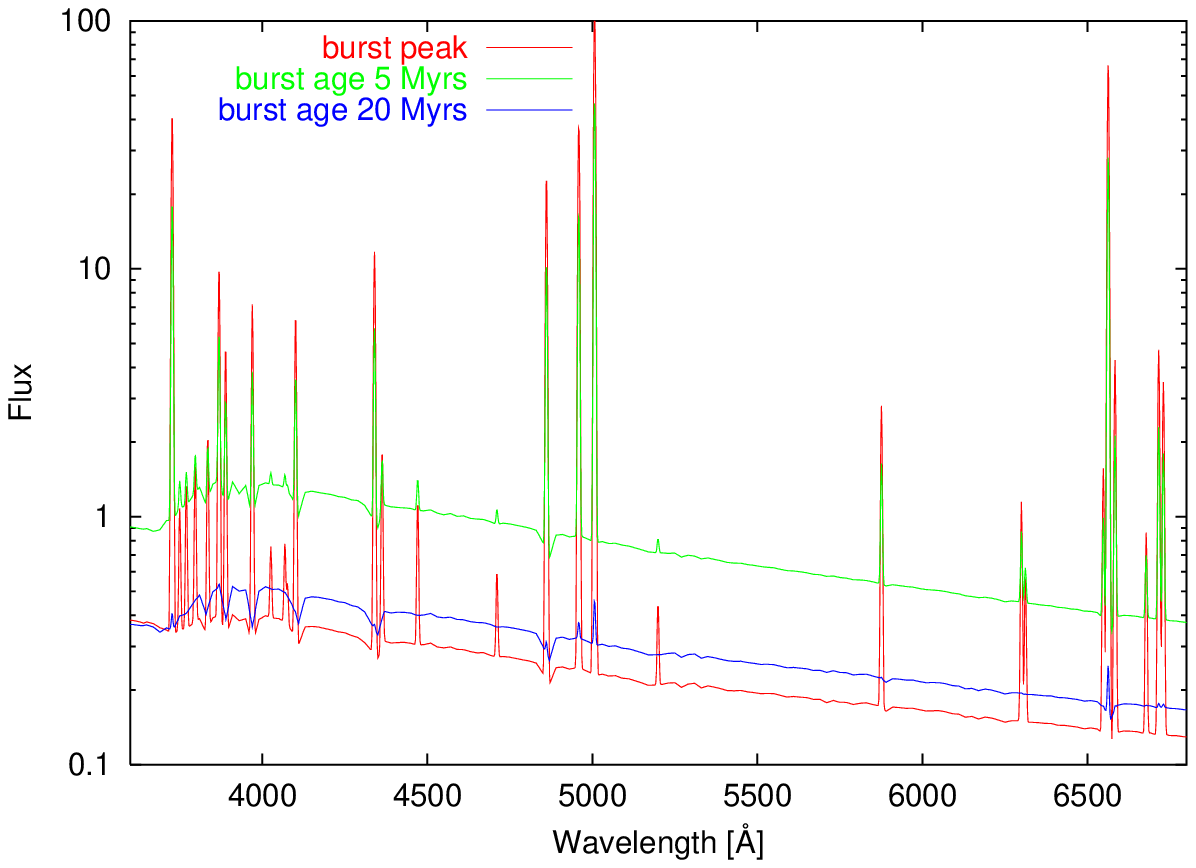,width=0.45\linewidth}
    \caption{Synthetic spectra for {\bf (a)} three different metallicities and 
      {\bf (b)} three different ages.}
    \label{fig:spec}
  \end{center}
\end{figure}

Three spectra of a strong starburst with different metallicities at
the same age are shown in Fig.~\ref{fig:spec}{\bf a}. The line fluxes
vary considerably due to changes in the metallicity of the gas and
different temperatures and luminosities of the young stars.  Three
more spectra show the time evolution during the burst for the model
with metallicity Z2 in Fig.~\ref{fig:spec}{\bf b}. The Balmer line
fluxes and the strength and slope of the continuum can be used to
select the best fit model for a given observation.

\section{First Results}\label{sec:FR}

\begin{figure}[t]
  \begin{center}
    \epsfig{file=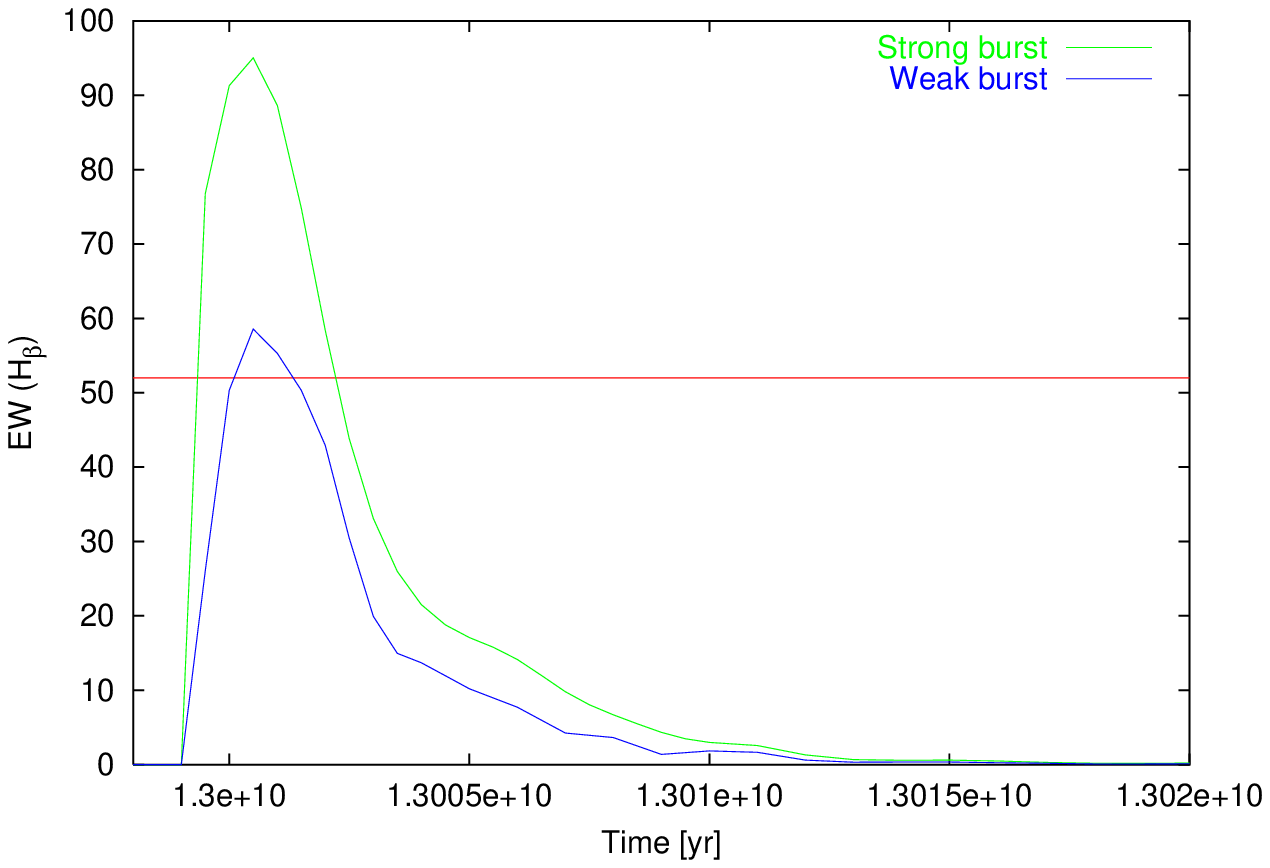,width=0.45\linewidth}
    \epsfig{file=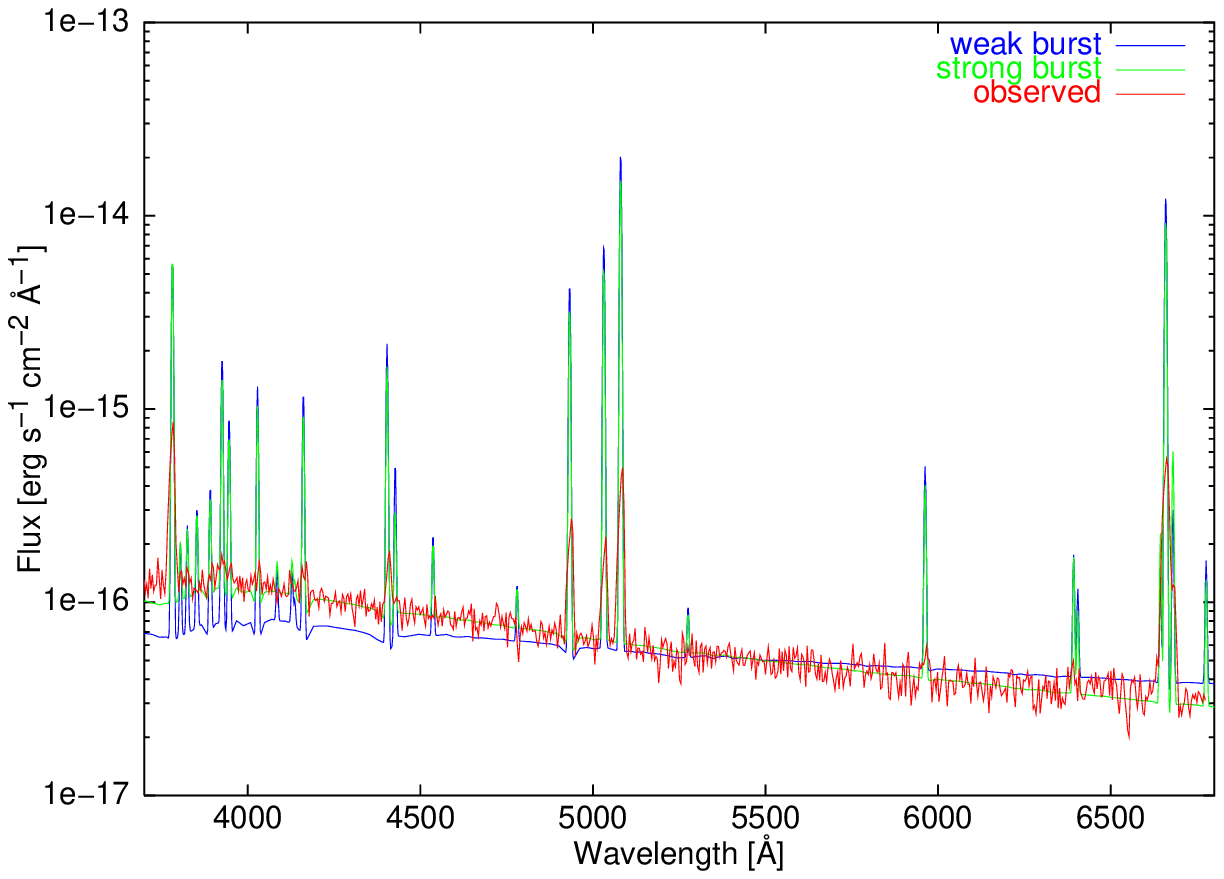,width=0.45\linewidth}
    \caption{{\bf (a)} The evolution of EW(H$_\beta$). 
                        The straight line is the observed value.
      {\bf (b)} Models with two different burst strengths at 
                 the allowed age compared with the observed spectrum.}
    \label{fig:specmodobs}
  \end{center}
\end{figure}

As an example of the procedure we use data of a typical TDG.  We first
use the broad-band colors to derive a good estimate of the burst
strength and a first guess of the burst age (as in
Fig.~\ref{fig:broad}).  The observed metal abundance will tell us
which metallicity to use for in the models.  We then compare the
observed equivalent width of H$_\beta$ with the equivalent width from
these models (two models are shown in Fig.~\ref{fig:specmodobs}{\bf
  a}). One of the possible ages for each model generally turns out to
be consistent with the age estimated from the broad-band photometry.

The dereddened observed spectrum of the same TDG plotted together with
the model spectra of weak and strong bursts preselected for the
correct burst age from their H$_\beta$ equivalent widths are given in
Fig.~\ref{fig:specmodobs}{\bf b}. We can finally select the correct
model using the slope of the continuum of the different models.

The results from the broad-band colors are confirmed in this case by
using the spectral information, a strong burst with $\sim$20\% young
stars and an age of 3 Myrs best explains the observed spectrum.

\section{Outlook}

The models shown here are only examples of what we plan to do. We will
compute a grid of spectral evolutionary models similar to the
photometric ones of \citet{WDF+00} with our updated input physics. We
also have to automate the procedure outlined in Sect.~\ref{sec:FR} to
compare the model grid with the observations of TDG candidates.
Finally we will apply this procedure to every TDG spectrum we have
(see Weilbacher \& Duc, this volume) to derive the burst parameters of
each object and to assess its current state and future evolution.

{\small {\bf Acknowledgement}. PMW is partially supported by the Deutsche
  Forschungsgemeinschaft (DFG Grant FR 916/6-1).}

\renewcommand{\refname}{{\small\bf References}}
\itemsep=0pt \parsep=0pt \parskip=0pt \labelsep=0pt
{\small
  \bibsep=0pt
  \bibliography{../../../PmW}
}

\end{document}